# Temperature-dependent refractive index of silicon and germanium

Bradley J. Frey[*], Douglas B. Leviton, Timothy J. Madison
NASA Goddard Space Flight Center, Greenbelt, MD  20771

## ABSTRACT

Silicon and germanium are perhaps the two most well-understood semiconductor materials in the context of solid state device technologies and more recently micromachining and nanotechnology.  Meanwhile, these two materials are also important in the field of infrared lens design.  Optical instruments designed for the wavelength range where these two materials are transmissive achieve best performance when cooled to cryogenic temperatures to enhance signal from the scene over instrument background radiation.  In order to enable high quality lens designs using silicon and germanium at cryogenic temperatures, we have measured the absolute refractive index of multiple prisms of these two materials using the Cryogenic, High-Accuracy Refraction Measuring System (CHARMS) at NASA's Goddard Space Flight Center, as a function of both wavelength and temperature.  For silicon, we report absolute refractive index and thermo-optic coefficient (dn/dT) at temperatures ranging from 20 to 300 K at wavelengths from 1.1 to 5.6 μm, while for germanium, we cover temperatures ranging from 20 to 300 K and wavelengths from 1.9 to 5.5 μm.  We compare our measurements with others in the literature and provide temperature-dependent Sellmeier coefficients based on our data to allow accurate interpolation of index to other wavelengths and temperatures.  Citing the wide variety of values for the refractive indices of these two materials found in the literature, we reiterate the importance of measuring the refractive index of a sample from the same batch of raw material from which final optical components are cut when absolute accuracy greater than $\pm 5 \times 10^{-3}$ is desired.

## 1.  INTRODUCTION

Historically, few accurate refractive index measurements of infrared materials have been made at cryogenic temperatures.  This has hampered the developments of many cryogenic infrared instruments, including the Infrared Array Camera (IRAC) for NASA's Spitzer Space Telescope, for which, for design purposes, index values for its lens materials were extrapolated from literature values both in wavelength and in temperature.  Such an approach leads to integration cycles which are far longer than anticipated, where best instrument performance is achieved by trial and error in many time-consuming and expensive iterations of instrument optical alignment.

The Cryogenic High Accuracy Refraction Measuring System (CHARMS) has been recently developed at GSFC to make such measurements in the absolute sense (in vacuum) down to temperatures as low as 15 K with unsurpassed accuracy using the method of minimum deviation refractometry.[1,2,3]  For low index materials with only modest dispersion such as synthetic fused silica, CHARMS can measure absolute index with an uncertainty in the sixth decimal place of index.  For materials with higher indices and high dispersion such as Si or Ge, CHARMS can measure absolute index with an uncertainty of about one part in the fourth decimal place of index.  Measurement and calibration methods used and recent facility improvements are discussed in previous papers.[4,5]

Prism pairs were purchased for this study of silicon and germanium by the James Webb Space Telescope's Near-Infrared Camera (JWST/NIRCam) and were fabricated from one boule of each respective material.  Therefore, this study is not primarily an interspecimen variability study for specimens from different material suppliers or from different lots of material from a given supplier.  Both silicon prisms were specified as "optical grade" silicon and both germanium prisms were specified as single-crystal germanium at the time of fabrication.  The purity level for all four prisms is specified at >99.999% by the material vendor.  Once again, no quantitative analysis of the relationship between crystal type or material purity and refractive index has been done during this study. The examination of such variability is being considered for future efforts.

---

[*] Brad.Frey@nasa.gov, phone 1-301-286-7787, FAX 1-301-286-0204

The apex angle of the prism for each material is designed so that the beam deviation angle for the highest index in the material's transparent range will equal the largest accessible deviation angle of the refractometer, 60°. Common prism dimensions are refracting face length and height of 38.1 mm and 28.6 mm, respectively. Nominal apex angles for the prisms measured in this study are: Si – 20.0° and Ge – 17.0°.

## 2. PRESENTATION OF MEASURED INDEX DATA

Detailed descriptions of our data acquisition and reduction processes are documented elsewhere[4], as are our calibration procedures[5]. In general we fit our raw measured data to a temperature dependent Sellmeier model of the form:

$$n^2(\lambda, T) - 1 = \sum_{i=1}^{m} \frac{S_i(T) \cdot \lambda^2}{\lambda^2 - \lambda_i^2(T)}$$

where $S_i$ are the strengths of the resonance features in the material at wavelengths $\lambda_i$. When dealing with a wavelength interval between wavelengths of physical resonances in the material, the summation may be approximated by only a few terms, m – typically three.[6] In such an approximation, resonance strengths $S_i$ and wavelengths $\lambda_i$ no longer have direct physical significance but are rather parameters used to generate an adequately accurate fit to empirical data. If these parameters are assumed to be functions of T, one can generate a temperature-dependent Sellmeier model for $n(\lambda,T)$.

Historically, this modeling approach has been employed with significant success for a variety of materials despite a rather serious sparsity of available index measurements – to cover a wide range of temperatures and wavelengths – upon which to base a model. One solution to the shortcoming of lack of measured index data has been to appeal to room temperature refractive index data at several wavelengths to anchor the model and then to extrapolate index values for other temperatures using accurate measurements of the thermo-optic coefficient dn/dT at those temperatures, which are much easier to make than accurate measurements of the index itself at exotic temperatures.[6] This is of course a potentially dangerous assumption, depending on the sample material in question and the required accuracy of refractive index knowledge.

Meanwhile, with CHARMS, we have made direct measurements of index densely sampled over a wide range of wavelengths and temperatures to produce a model with residuals on the order of the uncertainties in our raw index measurements. For our models, we have found that 4th order temperature dependences in all three terms in each of $S_i$ and $\lambda_i$ work adequately well, as also found previously in the literature.[6] The Sellmeier equation consequently becomes:

$$n^2(\lambda, T) - 1 = \sum_{i=1}^{3} \frac{S_i(T) \cdot \lambda^2}{\lambda^2 - \lambda_i^2(T)}$$

where,
These Sellmeier models are our best statistical representation of the measured data over the complete measured ranges of wavelength and temperature. All of the tabulated values for the refractive index of Si and Ge below have been

$$S_i(T) = \sum_{j=0}^{4} S_{ij} \cdot T^j$$

$$\lambda_i(T) = \sum_{j=0}^{4} \lambda_{ij} \cdot T^j$$

calculated using this Sellmeier model based on our measured data using the appropriate coefficients in Tables 5 and 10. Depending on the sample material measured, the residuals of the measured values compared to the fits can be as low as several parts in the sixth decimal place of index. This level of fit quality for data obtained using CHARMS generally pertains to low index materials with moderately low dispersion such as fused silica, LiF, and $BaF_2$. For high index materials such as Si and Ge, residuals are higher and in the range of $1 \times 10^{-4}$ to $1 \times 10^{-5}$ depending on the wavelength and temperature of interest.

## 2.1 Silicon

Absolute refractive indices of Si were measured over the 1.1 to 5.6 microns wavelength range and over the temperature range from 20 to 300 K for two test specimens which yielded equal indices to within our typical measurement uncertainty of +/- 1 x $10^{-4}$ (see Table 1). Indices are plotted in Figure 1, tabulated in Table 2 for selected temperatures and wavelengths. Spectral dispersion is plotted in Figure 2, tabulated in Table 3. Thermo-optic coefficient is plotted in Figure 3, tabulated in Table 4. Coefficients for the three term Sellmeier model with $4^{th}$ order temperature dependence are given in Table 5.

Table 1: Uncertainty in absolute refractive index measurements of silicon for selected wavelengths and temperatures

| wavelength | 30 K | 75 K | 100 K | 200 K | 295 K |
|---|---|---|---|---|---|
| 1.5 microns | 8.61E-05 | 1.35E-04 | 1.20E-04 | 9.16E-05 | 8.67E-05 |
| 3.0 microns | 5.00E-05 | 1.08E-04 | 9.14E-05 | 5.52E-05 | 4.55E-05 |
| 4.0 microns | 4.94E-05 | 1.07E-04 | 9.07E-05 | 5.44E-05 | 4.47E-05 |
| 5.0 microns | 5.08E-05 | 1.07E-04 | 9.00E-05 | 5.42E-05 | 4.46E-05 |

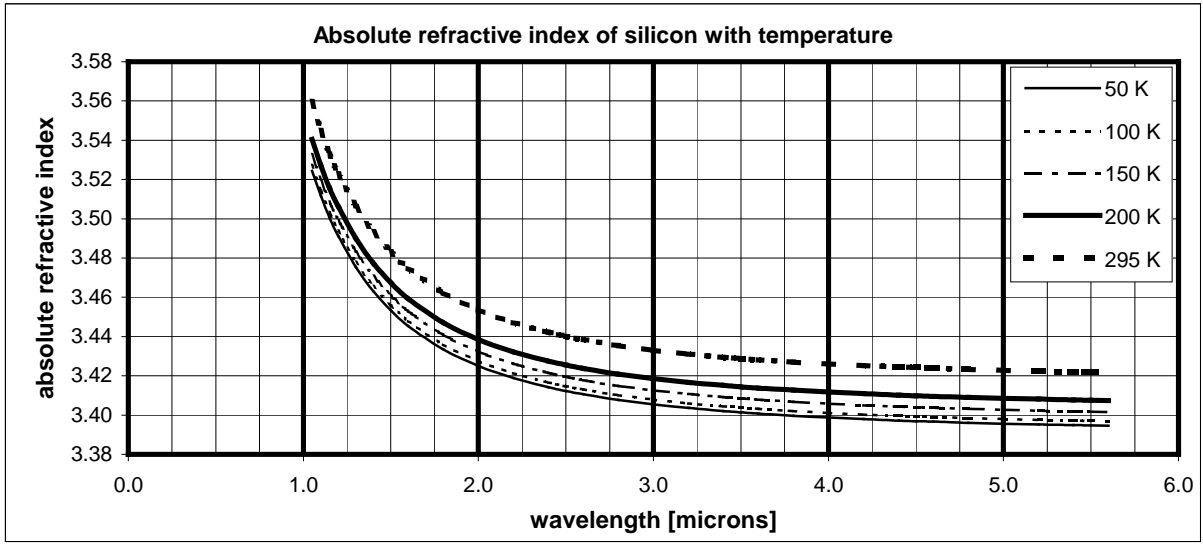

Figure 1: Measured absolute refractive index of silicon as a function of wavelength for selected temperatures

Table 2: Measured absolute refractive index of silicon for selected wavelengths and temperatures

| wavelength | 30 K | 40 K | 50 K | 60 K | 70 K | 80 K | 90 K | 100 K | 150 K | 200 K | 250 K | 295 K |
|---|---|---|---|---|---|---|---|---|---|---|---|---|
| 1.1 microns | 3.51113 | 3.51123 | 3.51146 | 3.51181 | 3.51229 | 3.51288 | 3.51358 | 3.51439 | 3.51979 | 3.52703 | 3.53555 | 3.54394 |
| 1.2 microns | 3.49071 | 3.49080 | 3.49103 | 3.49137 | 3.49183 | 3.49241 | 3.49309 | 3.49387 | 3.49910 | 3.50610 | 3.51437 | 3.52253 |
| 1.3 microns | 3.47508 | 3.47518 | 3.47540 | 3.47574 | 3.47619 | 3.47675 | 3.47741 | 3.47817 | 3.48327 | 3.49011 | 3.49818 | 3.50616 |
| 1.4 microns | 3.46285 | 3.46295 | 3.46317 | 3.46350 | 3.46394 | 3.46449 | 3.46514 | 3.46589 | 3.47089 | 3.47759 | 3.48551 | 3.49335 |
| 1.5 microns | 3.45309 | 3.45319 | 3.45340 | 3.45373 | 3.45417 | 3.45471 | 3.45535 | 3.45609 | 3.46100 | 3.46760 | 3.47540 | 3.48314 |
| 2.0 microns | 3.42478 | 3.42488 | 3.42508 | 3.42540 | 3.42582 | 3.42633 | 3.42695 | 3.42765 | 3.43234 | 3.43863 | 3.44609 | 3.45352 |
| 2.5 microns | 3.41196 | 3.41206 | 3.41226 | 3.41257 | 3.41298 | 3.41348 | 3.41408 | 3.41477 | 3.41936 | 3.42552 | 3.43283 | 3.44011 |
| 3.0 microns | 3.40509 | 3.40518 | 3.40538 | 3.40569 | 3.40609 | 3.40659 | 3.40718 | 3.40786 | 3.41240 | 3.41849 | 3.42572 | 3.43293 |
| 3.5 microns | 3.40100 | 3.40109 | 3.40128 | 3.40158 | 3.40198 | 3.40248 | 3.40307 | 3.40374 | 3.40826 | 3.41430 | 3.42148 | 3.42864 |
| 4.0 microns | 3.39837 | 3.39846 | 3.39866 | 3.39895 | 3.39935 | 3.39984 | 3.40042 | 3.40110 | 3.40560 | 3.41162 | 3.41877 | 3.42589 |
| 4.5 microns | 3.39661 | 3.39669 | 3.39688 | 3.39718 | 3.39757 | 3.39806 | 3.39864 | 3.39931 | 3.40381 | 3.40982 | 3.41694 | 3.42404 |
| 5.0 microns | 3.39537 | 3.39546 | 3.39564 | 3.39593 | 3.39632 | 3.39681 | 3.39739 | 3.39806 | 3.40256 | 3.40855 | 3.41566 | 3.42274 |
| 5.5 microns | 3.39449 | 3.39457 | 3.39475 | 3.39504 | 3.39542 | 3.39591 | 3.39649 | 3.39716 | 3.40166 | 3.40765 | 3.41474 | 3.42180 |

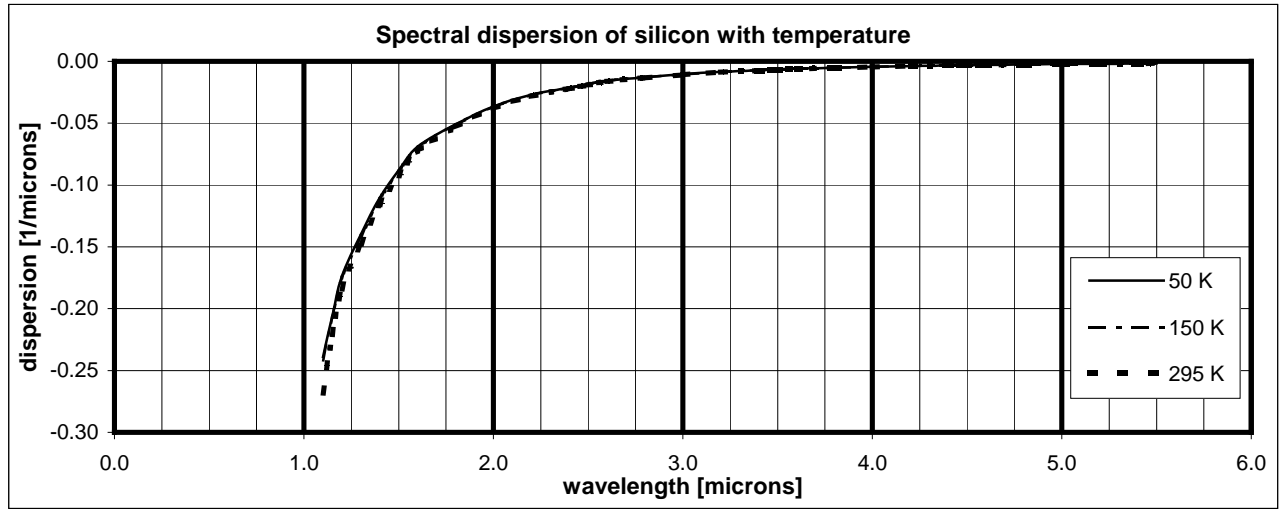

Figure 2: Measured spectral dispersion (dn/dλ) of silicon as a function of wavelength for selected temperatures

Table 3: Measured spectral dispersion (dn/dλ) of silicon for selected wavelengths and temperatures

| wavelength | 30 K | 40 K | 50 K | 60 K | 70 K | 80 K | 90 K | 100 K | 150 K | 200 K | 250 K | 295 K |
|---|---|---|---|---|---|---|---|---|---|---|---|---|
| 1.1 microns | -0.23895 | -0.23899 | -0.23908 | -0.23922 | -0.23942 | -0.23966 | -0.23996 | -0.24120 | -0.24191 | -0.24650 | -0.25586 | -0.26836 |
| 1.2 microns | -0.17503 | -0.17444 | -0.17402 | -0.17377 | -0.17369 | -0.17378 | -0.17403 | -0.17441 | -0.17548 | -0.17758 | -0.18065 | -0.18426 |
| 1.3 microns | -0.13879 | -0.13940 | -0.13991 | -0.14032 | -0.14063 | -0.14084 | -0.14095 | -0.14102 | -0.14222 | -0.14380 | -0.14583 | -0.14803 |
| 1.4 microns | -0.11024 | -0.11029 | -0.11034 | -0.11039 | -0.11046 | -0.11054 | -0.11062 | -0.11063 | -0.11156 | -0.11280 | -0.11429 | -0.11583 |
| 1.5 microns | -0.08795 | -0.08798 | -0.08802 | -0.08806 | -0.08812 | -0.08818 | -0.08825 | -0.08829 | -0.08893 | -0.08986 | -0.09103 | -0.09229 |
| 2.0 microns | -0.03646 | -0.03646 | -0.03647 | -0.03649 | -0.03651 | -0.03654 | -0.03659 | -0.03664 | -0.03689 | -0.03724 | -0.03769 | -0.03819 |
| 2.5 microns | -0.01814 | -0.01813 | -0.01814 | -0.01816 | -0.01818 | -0.01821 | -0.01826 | -0.01829 | -0.01843 | -0.01860 | -0.01878 | -0.01896 |
| 3.0 microns | -0.01036 | -0.01037 | -0.01037 | -0.01036 | -0.01036 | -0.01036 | -0.01035 | -0.01034 | -0.01044 | -0.01056 | -0.01071 | -0.01086 |
| 3.5 microns | -0.00648 | -0.00650 | -0.00651 | -0.00651 | -0.00649 | -0.00647 | -0.00644 | -0.00652 | -0.00658 | -0.00659 | -0.00665 | -0.00677 |
| 4.0 microns | -0.00437 | -0.00436 | -0.00436 | -0.00436 | -0.00437 | -0.00438 | -0.00440 | -0.00446 | -0.00444 | -0.00443 | -0.00445 | -0.00452 |
| 4.5 microns | -0.00299 | -0.00299 | -0.00300 | -0.00301 | -0.00302 | -0.00303 | -0.00305 | -0.00307 | -0.00306 | -0.00307 | -0.00310 | -0.00314 |
| 5.0 microns | -0.00210 | -0.00215 | -0.00218 | -0.00220 | -0.00222 | -0.00222 | -0.00221 | -0.00219 | -0.00222 | -0.00224 | -0.00226 | -0.00228 |
| 5.5 microns | -0.00155 | -0.00162 | -0.00167 | -0.00168 | -0.00167 | -0.00163 | -0.00155 | -0.00156 | -0.00171 | -0.00173 | -0.00170 | -0.00164 |

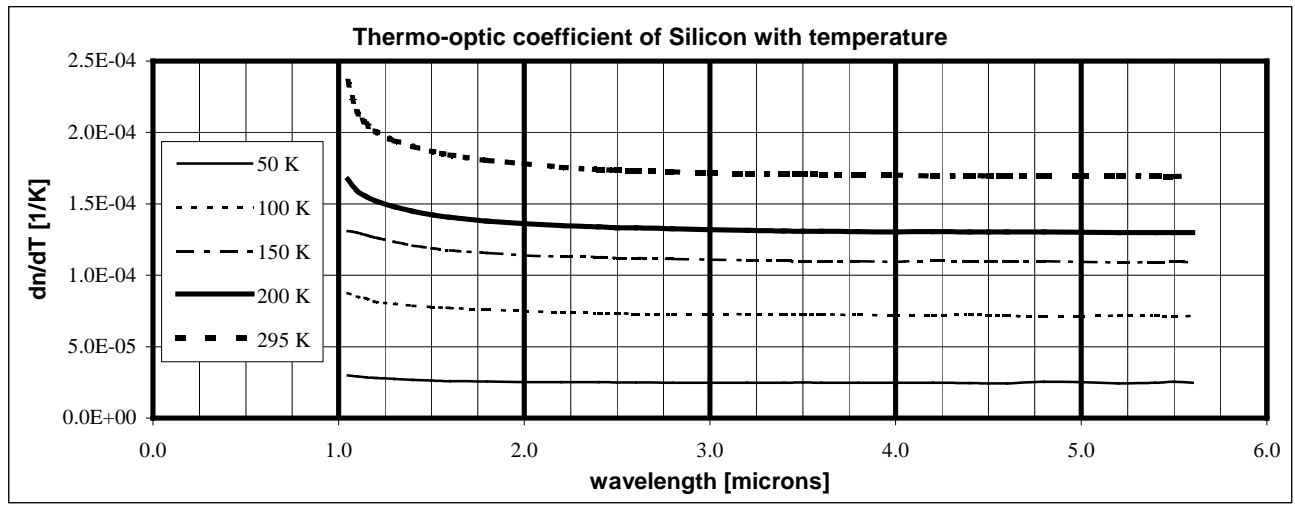

Figure 3: Measured thermo-optic coefficient (dn/dT) of silicon as a function of wavelength for selected temperatures

Table 4: Measured thermo-optic coefficient (dn/dT) of silicon for selected wavelengths and temperatures

| wavelength | 30 K | 40 K | 50 K | 60 K | 70 K | 80 K | 90 K | 100 K | 150 K | 200 K | 250 K | 295 K |
|---|---|---|---|---|---|---|---|---|---|---|---|---|
| 1.1 microns | 6.70E-06 | 1.79E-05 | 2.91E-05 | 4.04E-05 | 5.16E-05 | 6.28E-05 | 7.40E-05 | 8.52E-05 | 1.30E-04 | 1.59E-04 | 1.88E-04 | 2.15E-04 |
| 1.2 microns | 6.40E-06 | 1.72E-05 | 2.80E-05 | 3.86E-05 | 4.95E-05 | 6.08E-05 | 7.18E-05 | 8.14E-05 | 1.26E-04 | 1.52E-04 | 1.78E-04 | 2.01E-04 |
| 1.3 microns | 6.46E-06 | 1.70E-05 | 2.75E-05 | 3.80E-05 | 4.85E-05 | 5.90E-05 | 6.95E-05 | 8.00E-05 | 1.23E-04 | 1.48E-04 | 1.72E-04 | 1.94E-04 |
| 1.4 microns | 6.01E-06 | 1.64E-05 | 2.68E-05 | 3.72E-05 | 4.76E-05 | 5.80E-05 | 6.84E-05 | 7.88E-05 | 1.21E-04 | 1.45E-04 | 1.69E-04 | 1.90E-04 |
| 1.5 microns | 5.86E-06 | 1.61E-05 | 2.64E-05 | 3.67E-05 | 4.70E-05 | 5.72E-05 | 6.75E-05 | 7.78E-05 | 1.19E-04 | 1.42E-04 | 1.66E-04 | 1.87E-04 |
| 2.0 microns | 5.39E-06 | 1.53E-05 | 2.53E-05 | 3.52E-05 | 4.51E-05 | 5.51E-05 | 6.50E-05 | 7.50E-05 | 1.14E-04 | 1.36E-04 | 1.58E-04 | 1.78E-04 |
| 2.5 microns | 5.51E-06 | 1.52E-05 | 2.50E-05 | 3.47E-05 | 4.44E-05 | 5.41E-05 | 6.39E-05 | 7.36E-05 | 1.12E-04 | 1.33E-04 | 1.55E-04 | 1.74E-04 |
| 3.0 microns | 5.53E-06 | 1.51E-05 | 2.47E-05 | 3.44E-05 | 4.40E-05 | 5.36E-05 | 6.32E-05 | 7.28E-05 | 1.11E-04 | 1.32E-04 | 1.53E-04 | 1.71E-04 |
| 3.5 microns | 5.72E-06 | 1.53E-05 | 2.48E-05 | 3.44E-05 | 4.40E-05 | 5.35E-05 | 6.31E-05 | 7.27E-05 | 1.10E-04 | 1.31E-04 | 1.52E-04 | 1.71E-04 |
| 4.0 microns | 5.90E-06 | 1.53E-05 | 2.48E-05 | 3.42E-05 | 4.36E-05 | 5.30E-05 | 6.25E-05 | 7.19E-05 | 1.09E-04 | 1.30E-04 | 1.51E-04 | 1.70E-04 |
| 4.5 microns | 5.20E-06 | 1.47E-05 | 2.43E-05 | 3.38E-05 | 4.34E-05 | 5.29E-05 | 6.25E-05 | 7.20E-05 | 1.10E-04 | 1.30E-04 | 1.51E-04 | 1.70E-04 |
| 5.0 microns | 6.79E-06 | 1.60E-05 | 2.52E-05 | 3.44E-05 | 4.36E-05 | 5.28E-05 | 6.20E-05 | 7.13E-05 | 1.09E-04 | 1.30E-04 | 1.51E-04 | 1.69E-04 |
| 5.5 microns | 7.11E-06 | 1.62E-05 | 2.54E-05 | 3.45E-05 | 4.36E-05 | 5.28E-05 | 6.19E-05 | 7.10E-05 | 1.09E-04 | 1.30E-04 | 1.51E-04 | 1.69E-04 |

Table 5: Coefficients for the temperature-dependent Sellmeier fit of the refractive index of Si measured by CHARMS. Average absolute residual of the fit from the measured data is $1 \times 10^{-4}$; the measurement uncertainty for Si is listed in Table 1.

| **Coefficients for the temperature dependent Sellmeier equation for Si** | | | | | | |
|---|---|---|---|---|---|---|
| **20 K <= T <= 300 K; 1.1 microns <= $\lambda$ <= 5.6 microns** | | | | | | |
| | $S_1$ | $S_2$ | $S_3$ | $\lambda_1$ | $\lambda_2$ | $\lambda_3$ |
| **Constant term** | 10.4907 | -1346.61 | 4.42827E+07 | 0.299713 | -3.51710E+03 | 1.71400E+06 |
| **T term** | -2.08020E-04 | 29.1664 | -1.76213E+06 | -1.14234E-05 | 42.3892 | -1.44984E+05 |
| **$T^2$ term** | 4.21694E-06 | -0.278724 | -7.61575E+04 | 1.67134E-07 | -0.357957 | -6.90744E+03 |
| **$T^3$ term** | -5.82298E-09 | 1.05939E-03 | 678.414 | -2.51049E-10 | 1.17504E-03 | -39.3699 |
| **$T^4$ term** | 3.44688E-12 | -1.35089E-06 | 103.243 | 2.32484E-14 | -1.13212E-06 | 23.5770 |

## 2.2 Germanium

Absolute refractive indices of Ge were measured over the 1.9 to 5.5 microns wavelength range and over a range of temperatures from 20 to 300 K for two test specimens which, as with Si, yielded equal indices to within our typical measurement uncertainty of +/-1 x $10^{-4}$ (see Table 6). Indices are plotted in Figure 4, tabulated in Table 7 for selected temperatures and wavelengths. Spectral dispersion is plotted in Figure 5, tabulated in Table 8. Thermo-optic coefficient is plotted in Figure 6, tabulated in Table 9. Coefficients for the three term Sellmeier model with $4^{th}$ order temperature dependence are given in Table 10.

Table 6: Uncertainty in absolute refractive index measurements of germanium for selected wavelengths and temperatures

| wavelength | 30 K | 75 K | 100 K | 200 K | 295 K |
|---|---|---|---|---|---|
| 2.0 microns | 2.09E-04 | 2.33E-04 | 1.81E-04 | 1.49E-04 | 1.21E-04 |
| 3.0 microns | 1.68E-04 | 2.05E-04 | 1.48E-04 | 1.11E-04 | 6.97E-05 |
| 4.0 microns | 1.75E-04 | 1.93E-04 | 1.44E-04 | 1.06E-04 | 6.52E-05 |
| 5.0 microns | 1.74E-04 | 1.93E-04 | 1.41E-04 | 1.04E-04 | 6.41E-05 |

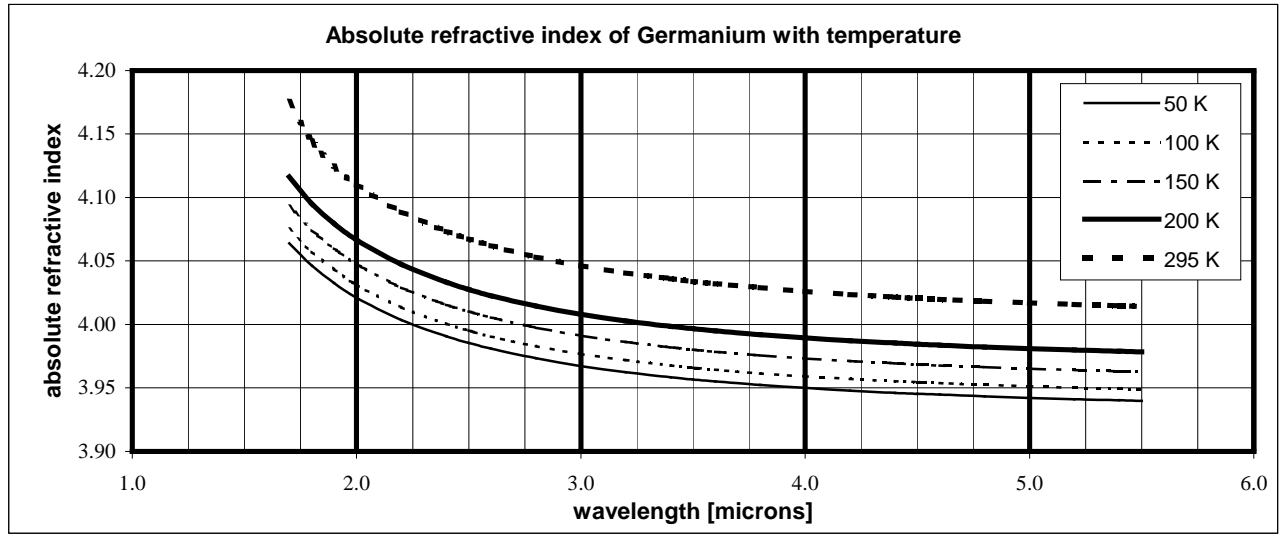

Figure 4: Measured absolute refractive index of germanium as a function of wavelength for selected temperatures

Table 7: Measured absolute refractive index of germanium for selected wavelengths and temperatures

| wavelength | 30 K | 40 K | 50 K | 60 K | 70 K | 80 K | 90 K | 100 K | 150 K | 200 K | 250 K | 295 K |
|---|---|---|---|---|---|---|---|---|---|---|---|---|
| 1.8 microns | 4.04450 | 4.04529 | 4.04647 | 4.04801 | 4.04990 | 4.05210 | 4.05459 | 4.05735 | 4.07435 | 4.09518 | 4.11887 | 4.14224 |
| 2.0 microns | 4.01922 | 4.01997 | 4.02109 | 4.02255 | 4.02433 | 4.02641 | 4.02877 | 4.03136 | 4.04733 | 4.06673 | 4.08853 | 4.10973 |
| 2.5 microns | 3.98368 | 3.98439 | 3.98544 | 3.98681 | 3.98847 | 3.99041 | 3.99260 | 3.99502 | 4.00981 | 4.02770 | 4.04763 | 4.06693 |
| 3.0 microns | 3.96562 | 3.96630 | 3.96732 | 3.96865 | 3.97026 | 3.97214 | 3.97425 | 3.97659 | 3.99088 | 4.00812 | 4.02728 | 4.04579 |
| 3.5 microns | 3.95506 | 3.95574 | 3.95675 | 3.95805 | 3.95963 | 3.96147 | 3.96355 | 3.96584 | 3.97985 | 3.99673 | 4.01548 | 4.03355 |
| 4.0 microns | 3.94833 | 3.94901 | 3.95001 | 3.95130 | 3.95287 | 3.95468 | 3.95674 | 3.95900 | 3.97285 | 3.98949 | 4.00798 | 4.02577 |
| 4.5 microns | 3.94377 | 3.94445 | 3.94544 | 3.94673 | 3.94829 | 3.95009 | 3.95213 | 3.95438 | 3.96812 | 3.98459 | 4.00291 | 4.02051 |
| 5.0 microns | 3.94052 | 3.94121 | 3.94221 | 3.94349 | 3.94505 | 3.94685 | 3.94887 | 3.95111 | 3.96478 | 3.98113 | 3.99932 | 4.01678 |
| 5.5 microns | 3.93813 | 3.93883 | 3.93983 | 3.94112 | 3.94267 | 3.94447 | 3.94649 | 3.94872 | 3.96235 | 3.97858 | 3.99668 | 4.01404 |

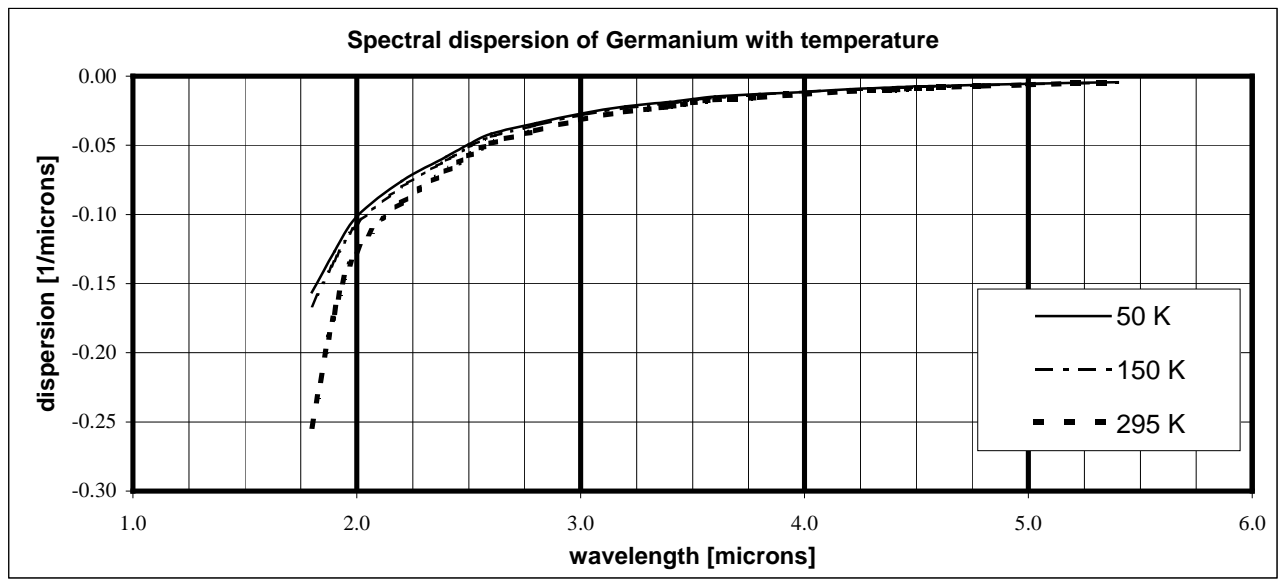

Figure 5: Measured spectral dispersion (dn/dλ) of germanium as a function of wavelength for selected temperatures

Table 8: Measured spectral dispersion (dn/dλ) of germanium for selected wavelengths and temperatures

| wavelength | 30 K | 40 K | 50 K | 60 K | 70 K | 80 K | 90 K | 100 K | 150 K | 200 K | 250 K | 295 K |
|---|---|---|---|---|---|---|---|---|---|---|---|---|
| 1.8 microns | -0.15561 | -0.15582 | -0.15624 | -0.15688 | -0.15773 | -0.16005 | -0.16057 | -0.16032 | -0.16640 | -0.18472 | -0.21528 | -0.25326 |
| 2.0 microns | -0.10125 | -0.10133 | -0.10151 | -0.10178 | -0.10214 | -0.10287 | -0.10354 | -0.10400 | -0.10730 | -0.11226 | -0.11889 | -0.12628 |
| 2.5 microns | -0.04893 | -0.04906 | -0.04913 | -0.04914 | -0.04908 | -0.04896 | -0.04926 | -0.04956 | -0.05120 | -0.05310 | -0.05525 | -0.05741 |
| 3.0 microns | -0.02701 | -0.02714 | -0.02722 | -0.02724 | -0.02721 | -0.02723 | -0.02744 | -0.02756 | -0.02828 | -0.02919 | -0.03030 | -0.03147 |
| 3.5 microns | -0.01659 | -0.01664 | -0.01666 | -0.01663 | -0.01658 | -0.01659 | -0.01677 | -0.01685 | -0.01730 | -0.01786 | -0.01852 | -0.01920 |
| 4.0 microns | -0.01121 | -0.01135 | -0.01141 | -0.01139 | -0.01129 | -0.01104 | -0.01100 | -0.01104 | -0.01135 | -0.01184 | -0.01248 | -0.01321 |
| 4.5 microns | -0.00739 | -0.00742 | -0.00747 | -0.00752 | -0.00758 | -0.00746 | -0.00736 | -0.00745 | -0.00787 | -0.00819 | -0.00844 | -0.00858 |
| 5.0 microns | -0.00554 | -0.00556 | -0.00556 | -0.00554 | -0.00551 | -0.00552 | -0.00562 | -0.00565 | -0.00583 | -0.00600 | -0.00615 | -0.00628 |

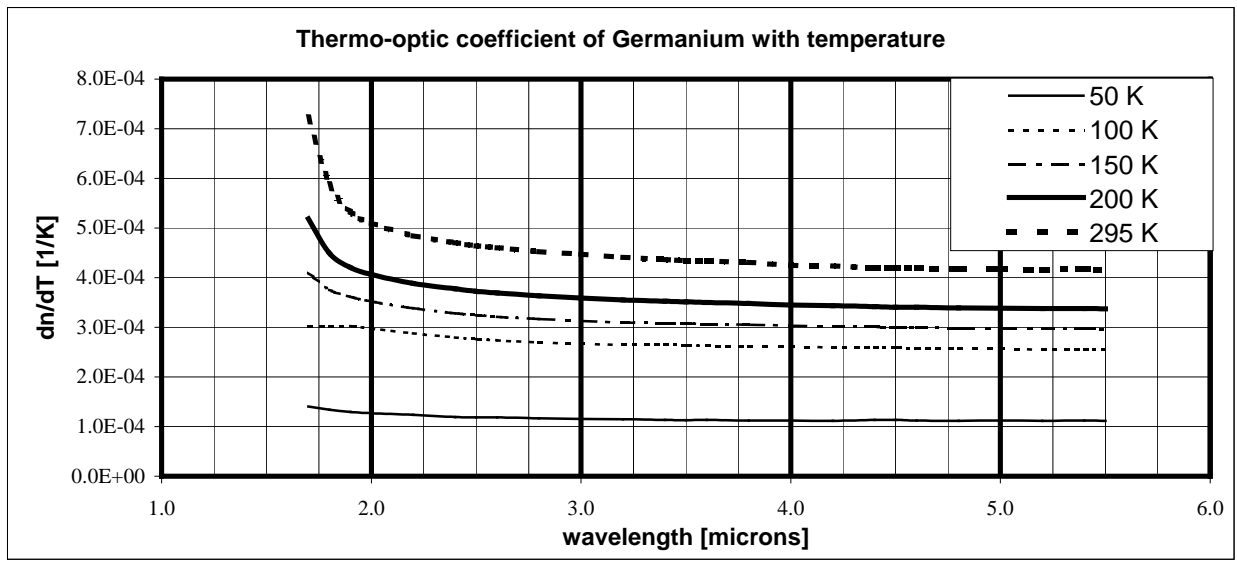

Figure 6: Measured thermo-optic coefficient (dn/dT) of germanium as a function of wavelength for selected temperatures

Table 9: Measured thermo-optic coefficient (dn/dT) of germanium for selected wavelengths and temperatures

| wavelength | 30 K | 40 K | 50 K | 60 K | 70 K | 80 K | 90 K | 100 K | 150 K | 200 K | 250 K | 295 K |
|---|---|---|---|---|---|---|---|---|---|---|---|---|
| 1.8 microns | 6.43E-05 | 9.92E-05 | 1.34E-04 | 1.69E-04 | 2.04E-04 | 2.56E-04 | 2.87E-04 | 3.02E-04 | 3.76E-04 | 4.49E-04 | 5.23E-04 | 5.89E-04 |
| 2.0 microns | 6.10E-05 | 9.39E-05 | 1.27E-04 | 1.60E-04 | 1.93E-04 | 2.51E-04 | 2.86E-04 | 2.97E-04 | 3.52E-04 | 4.06E-04 | 4.61E-04 | 5.10E-04 |
| 2.5 microns | 5.76E-05 | 8.82E-05 | 1.19E-04 | 1.50E-04 | 1.80E-04 | 2.34E-04 | 2.67E-04 | 2.76E-04 | 3.25E-04 | 3.73E-04 | 4.21E-04 | 4.64E-04 |
| 3.0 microns | 5.19E-05 | 8.37E-05 | 1.16E-04 | 1.47E-04 | 1.79E-04 | 2.30E-04 | 2.57E-04 | 2.66E-04 | 3.13E-04 | 3.59E-04 | 4.06E-04 | 4.47E-04 |
| 3.5 microns | 5.43E-05 | 8.36E-05 | 1.13E-04 | 1.42E-04 | 1.72E-04 | 2.24E-04 | 2.55E-04 | 2.64E-04 | 3.08E-04 | 3.51E-04 | 3.95E-04 | 4.34E-04 |
| 4.0 microns | 5.48E-05 | 8.34E-05 | 1.12E-04 | 1.41E-04 | 1.69E-04 | 2.21E-04 | 2.53E-04 | 2.61E-04 | 3.03E-04 | 3.45E-04 | 3.87E-04 | 4.25E-04 |
| 4.5 microns | 5.49E-05 | 8.41E-05 | 1.13E-04 | 1.42E-04 | 1.72E-04 | 2.21E-04 | 2.50E-04 | 2.59E-04 | 3.00E-04 | 3.41E-04 | 3.82E-04 | 4.19E-04 |
| 5.0 microns | 5.44E-05 | 8.32E-05 | 1.12E-04 | 1.41E-04 | 1.69E-04 | 2.19E-04 | 2.48E-04 | 2.56E-04 | 2.97E-04 | 3.38E-04 | 3.80E-04 | 4.17E-04 |
| 5.5 microns | 5.56E-05 | 8.35E-05 | 1.11E-04 | 1.39E-04 | 1.67E-04 | 2.17E-04 | 2.47E-04 | 2.55E-04 | 2.96E-04 | 3.37E-04 | 3.78E-04 | 4.16E-04 |

Table 10: Coefficients for the temperature-dependent Sellmeier fit of the refractive index of Ge measured by CHARMS. Average absolute residual of the fit from the measured data is 1 x 10$^{-4}$ and the measurement uncertainty for Ge is listed in Table 6.

| Coefficients for the temperature dependent Sellmeier equation for Ge 20 K <= T <= 300 K; 1.9 microns <= λ <= 5.5 microns | | | | | | |
|---|---|---|---|---|---|---|
| | $S_1$ | $S_2$ | $S_3$ | $\lambda_1$ | $\lambda_2$ | $\lambda_3$ |
| Constant term | 13.9723 | 0.452096 | 751.447 | 0.386367 | 1.08843 | -2893.19 |
| T term | 2.52809E-03 | -3.09197E-03 | -14.2843 | 2.01871E-04 | 1.16510E-03 | -0.967948 |
| T$^2$ term | -5.02195E-06 | 2.16895E-05 | -0.238093 | -5.93448E-07 | -4.97284E-06 | -0.527016 |
| T$^3$ term | 2.22604E-08 | -6.02290E-08 | 2.96047E-03 | -2.27923E-10 | 1.12357E-08 | 6.49364E-03 |
| T$^4$ term | -4.86238E-12 | 4.12038E-11 | -7.73454E-06 | 5.37423E-12 | 9.40201E-12 | -1.95162E-05 |

## 3. COMPARISON WITH LITERATURE VALUES

There have been many efforts in the past to measure the refractive indices of silicon[7,8,9,10,11,12] and germanium[7,10,12,13,14]. Comparisons between the present work and these literature values can be found in Figures 7 and 8. The references mentioned above are by no means an exhaustive compilation, but rather serve to provide a sample of the measurement values published using prismatic measurement techniques (i.e. minimum deviation method, modified minimum deviation method, etc.). Most investigators list uncertainties in their respective measurements somewhere in the range of ±1 x 10$^{-3}$ to ±1 x 10$^{-4}$, the present work included (see Tables 1 and 6). As other researchers have concluded, even though the range of measured values in the literature is greater than the listed uncertainties amongst the various authors, we believe our measurements are accurate to the level stated in the above tables. The differences between the reported values may in fact be real if there are significant interspecimen differences in refractive index due to sample material purity or other factors. Unfortunately a quantitative measure of these differences in physical properties of the measured samples has not been done, nor have different researchers measured the same sample prisms as far as we can tell. This conclusion only reinforces the philosophy that — depending on the accuracy requirements for a given optical design — a sample prism from the same batch of material as the final optical components should be fabricated and measured to insure to appropriate index of refraction values are used.

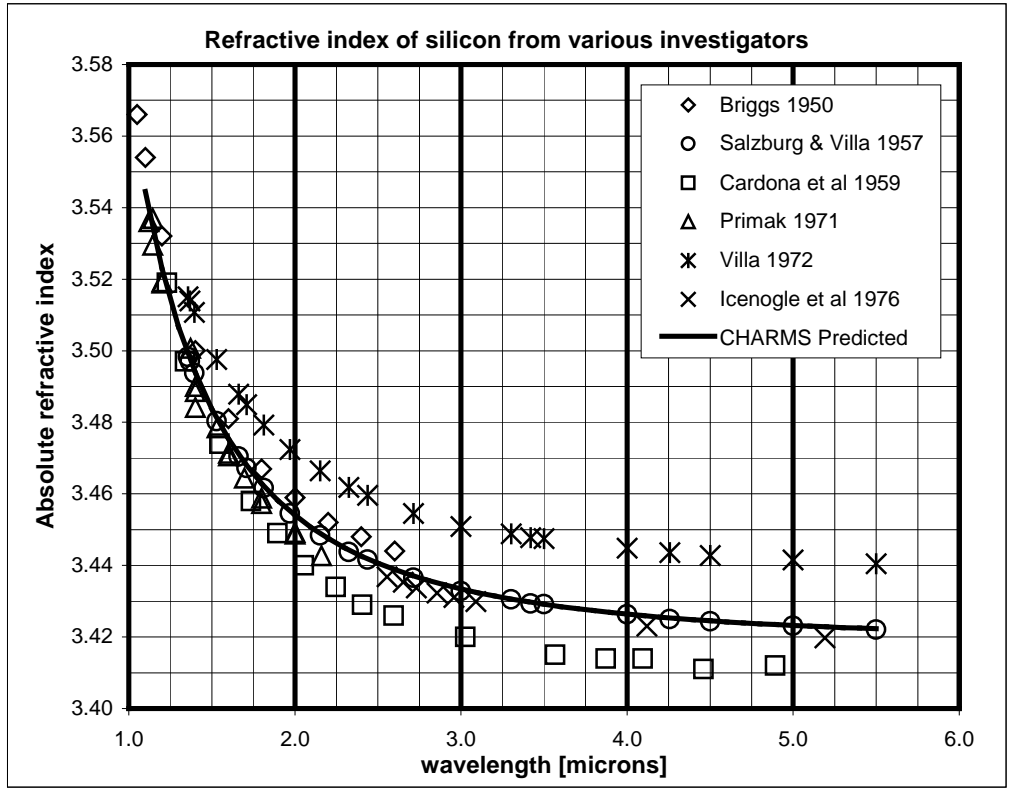

Figure 7: Comparison of the literature values for the absolute refractive index of silicon at 298K to the values measured by CHARMS.

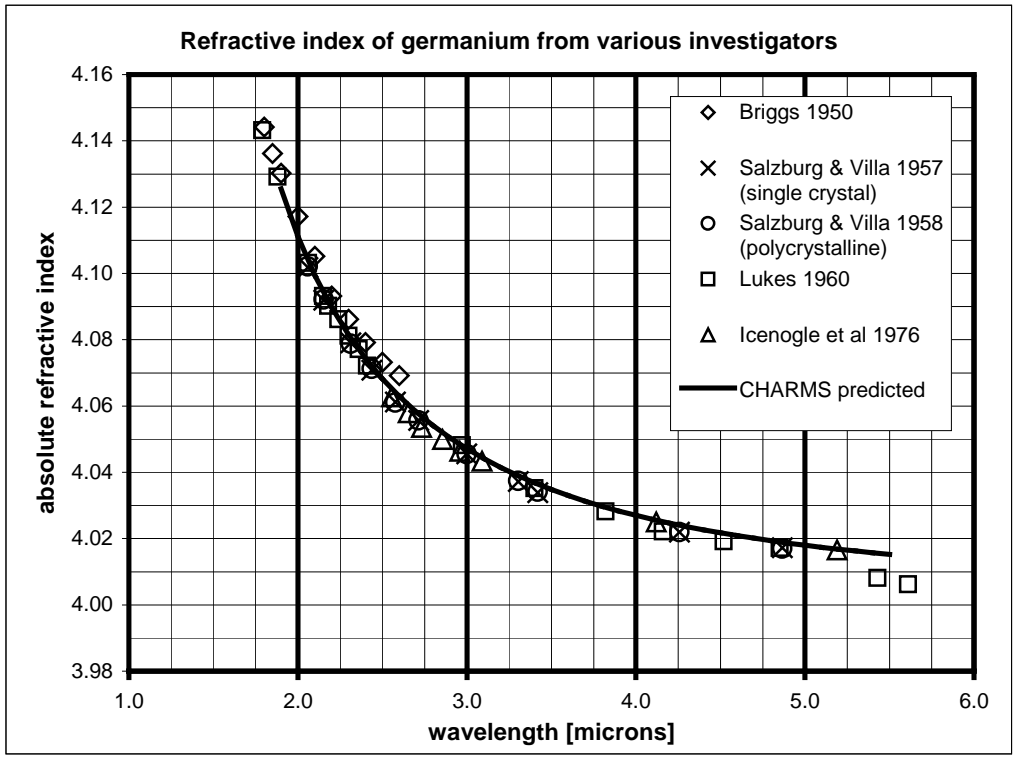

Figure 8: Comparison of the literature values for the absolute refractive index
of germanium at 298K to the values measured by CHARMS.

## 4. CONCLUSION

Using CHARMS, we have directly measured the absolute refractive indices of the infrared materials silicon and germanium, from their near-infrared cutoff to 5.6μm, and from room temperature down to as low as 20 K. Their spectral dispersion and thermo-optic coefficient were also examined and temperature dependent Sellmeier models were generated based on the measured refractive index data with residuals on the order of the uncertainty in the respective measurements. While our measurement uncertainty was smaller than discrepancies with other published values in the literature, we believe the differences can be at least partially attributed to interspecimen variability and variations in specimen purity. The conclusion has been drawn that for applications requiring accurate knowledge of the refractive index of their optical components, the measurement of a prism cut from the same batch of material as those optical components should be part of the optical design process.

## ACKNOWLEDGEMENTS

The authors would like to thank the JWST/NIRCam project for supplying the sample prisms and financial support for this study.

## REFERENCES

1. D.B. Leviton, B.J. Frey, "Design of a cryogenic, high accuracy, absolute prism refractometer for infrared through far ultraviolet optical materials," *Proc. SPIE* **4842**, 259-269 (2003)
2. B.J. Frey, R. Henry, D.B. Leviton, M. Quijada, "Cryogenic high-accuracy absolute prism refractometer for infrared through far-ultra-violet optical materials: implementation and initial results," *Proc. SPIE* **5172**, 119-129 (2003)
3. D.B. Leviton, B.J. Frey, "Cryogenic, High-Accuracy, Refraction Measuring System – a new facility for cryogenic infrared through far-ultraviolet refractive index measurements," *Proc. SPIE*, **5494**. 492-504 (2004)
4. B.J. Frey, D.B. Leviton, "Automation, operation, and data analysis in the cryogenic, high accuracy, refraction measuring system (CHARMS)," *Proc. SPIE*, **5904**, 59040P (2005)
5. D.B. Leviton, B.J. Frey, T.Kvamme, "High accuracy, absolute, cryogenic refractive index measurements of infrared lens materials for JWST NIRCam using CHARMS," *Proc. SPIE*, **5904**, 59040O (2005)
6. W.J. Tropf, "Temperature-dependent refractive index models for $BaF_2$, $CaF_2$, $MgF_2$, $SrF_2$, LiF, NaF, KCl, ZnS, and ZnSe," *Optical Engineering*, **34**(5), pp. 1369-1373, May 1995
7. H.B. Briggs, "Optical effects in bulk silicon and germanium," *Phy. Rev*. **77**, 287 (1950)
8. M. Cardona, W. Paul, and H. Brooks, "Dielectric constant of germanium and silicon as a function of volume," *Phys. Chem Solids*, **8**, 204-6 (1959)
9. W. Primak, "Refractive index of silicon," *Applied Optics*, **10**(4), 759-63 (1971)
10. C.D. Salzburg and J.J. Villa, "Infrared refractive indexes of silicon germanium and modified selenium glass," *J. Opt. Soc. Am,* **47**(3), 244-6 (1957)
11. J.J. Villa, "Additional data on the refractive index of silicon," *Applied Optics*, **11**(9) 2102-3 (1972)
12. H.W. Icenogle, B.C Platt, W.L. Wolfe, "Refractive indexes and temperature coefficients of germanium and silicon," *Applied Optics*, **15**(10), 2348-51 (1976)
13. C.D. Salzburg and J.J. Villa, "Index of refraction of germanium," *J. Opt. Soc. Am,* **48**(8), 579 (1958)
14. F. Lukes, "The temperature dependence of the refractive index of germanium," *Czech, J. Phys.,* **10**(10), 742-8 (1960)